\documentclass[aps,prl,twocolumn,showpacs,amssymb, footinbib]{revtex4-1}
\usepackage{amsfonts,amsmath,amssymb,amsthm,bbm,hyperref}
 \usepackage{eurosym}

\usepackage{color,psfrag}
\usepackage[dvips]{graphicx}

\usepackage[normalem]{ulem}

\newcommand{\twopoint}[1]{\left\langle #1 \right\rangle}

\newcommand{\m}{\mbox{}}

\newcommand{\be}{\begin{equation}}
\newcommand{\ee}{\end{equation}}
\newcommand{\ba}{\begin{eqnarray}}
\newcommand{\ea}{\end{eqnarray}}

\begin{document}

\title{A quantum reduction to spherical symmetry in loop quantum gravity}

\author{N. Bodendorfer}
\email[]{norbert.bodendorfer@fuw.edu.pl} 

\author{J. Lewandowski}
\email[]{jerzy.lewandowski@fuw.edu.pl}

\author{J. \'Swie\.zewski}
\email[]{swiezew@fuw.edu.pl}

\affiliation{Faculty of Physics, University of Warsaw, Pasteura 5, 02-093, Warsaw, Poland}

\date{{\small \today}}

\begin{abstract}

Based on a recent purely geometric construction of observables for the spatial diffeomorphism constraint, we propose two distinct quantum reductions to spherical symmetry within full $3+1$-dimensional loop quantum gravity. 
The construction of observables corresponds to using the radial gauge for the spatial metric and allows to identify rotations around a central observer as unitary transformations in the quantum theory. Group averaging over these rotations yields our first proposal for spherical symmetry. Hamiltonians of the full theory with angle-independent lapse preserve this spherically symmetric subsector of the full Hilbert space. 
A second proposal consists in implementing the vanishing of a certain vector field in spherical symmetry as a constraint on the full Hilbert space, leading to a close analogue of diffeomorphisms invariant states. While this second set of spherically symmetric states does not allow for using the full Hamiltonian, it is naturally suited to implement the spherically symmetric midisuperspace Hamiltonian, as an operator in the full theory, on it.
Due to the canonical structure of the reduced variables, the holonomy-flux algebra behaves effectively as a one parameter family of $2+1$-dimensional algebras along the radial coordinate, leading to a diagonal non-vanishing volume operator on $3$-valent vertices. The quantum dynamics thus becomes tractable, including scenarios like spherically symmetric dust collapse. 

\end{abstract}


\maketitle

\noindent {\bf Introduction} \\ 
Loop quantum gravity \cite{RovelliQuantumGravity, ThiemannModernCanonicalQuantum} as a whole has matured into a serious candidate theory for quantum gravity in recent years. Progress has been especially strong in the areas of computing black hole entropy \cite{AshtekarQuantumGeometryOf, EngleBlackHoleEntropy, BI} and studying quantisations of mini-\cite{AshtekarQuantumGravityExtension} or midi-\cite{GambiniLoopQuantizationOf}-superspace models using techniques from the full theory. On the other hand, it has been notoriously hard to extract physics from computations directly in the full $3+1$-dimensional theory. In order to avoid the ``problem of time'' associated with the underlying diffeomorphism invariance of general relativity, deparametrisation \cite{GornickaHamiltonianTheoryOf, BrownDustAsStandard} has been introduced within loop quantum gravity \cite{GieselAQG4, DomagalaGravityQuantizedLoop, HusainTimeAndA} in order to obtain a true Hamiltonian evolution. 
A certain form of deparametrisation, however, always puts restrictions on the physical situation that one can describe, due to a, in general, finite range of physical coordinates. It is therefore desirable to have different deparametrisation techniques at one's disposal, tailored to different interesting physical problems. Furthermore, deparametrisation can in principle significantly alter the canonical structure, leading to different quantisation variables. While this does not happen for the standard example of dust \cite{BrownDustAsStandard}, or only in a mild form of a possible rescaling for scalar fields \cite{BSTI} due to a Higgs-like ``absorption'' of matter degrees of freedom, we will encounter a more severe change of canonical structure due to a purely geometric deparametrisation in this article. As a direct consequence of this deparametrisation, we will obtain a family of holonomy-flux algebras labelled by the radial coordinate, each behaving effectively two-dimensional. Thus, we can use spin networks with three-valent vertices on which the volume operator is diagonal. The quantum dynamics thus becomes a lot more tractable than in the usual case. 
Also, the physical coordinate system introduced via this deparametrisation is ideally suited for introducing a quantum reduction to spherical symmetry. In order to simplify the presentation in this letter, we will gloss over some technical details which are addressed in our companion papers \cite{BLSII, BLSIII}.\\

\noindent {\bf The radial gauge} \\
Recently, a purely geometric construction of observables with respect to the spatial diffeomorphism constraint has been given \cite{DuchObservablesForGeneral}, based on a physical coordinate system introduced by spatial geodesics outgoing from a central point $\sigma_0$. A point in the spatial slice $\Sigma$ is uniquely defined via the exponential map $x^I \mapsto \exp_{\sigma_0} (x^I e^i_I)$, where $x^I$, $I=1,2,3$ are coordinates in an internal space and $e^i_I$ is a frame at $\sigma_0$ depending on a choice of a fiducial frame $e^i_{0I}$ at $\sigma_0$ as well as the spatial metric $q_{ij}$. $i,j = 1,2,3$ are local tensor indices on $\Sigma$. It is convenient to switch to spherical coordinates $a,b = r, A, B$ in the internal space; $r = \sqrt{x^I x_I}$ being the radial coordinate, and $A,B$ are angular coordinates, often abbreviated by $\theta$. 
Once the tensor indices are adapted to the induced coordinate system on $\Sigma$ ($i,j \rightarrow a,b$), the observable $Q_{ab}$ corresponding to the spatial metric satisfies $Q_{ra} = \delta_{ra}$ \footnote{In certain situations, e.g., to describe a Schwarzschild black hole in Gullstrand-Painlev\'e coordinates, it is advantageous to invert this construction in such a way that proper distance is counted from spatial infinity, see \cite{BLSII} for details.}. 

This construction of observables has an analogue in terms of a gauge fixing of the spatial diffeomorphism constraint, which we will employ in this paper. It is similar to the {\em radial gauge} employed in numerical relativity, see e.g. \cite{MathewsRelativisticNumericalHydrodynamics}, which is why we adopt this terminology. We start with the ADM phase space, subject to the Poisson bracket $\left\{q_{ij}(\sigma), p^{kl}(\sigma')\right\} = \delta_{(i}^k \delta_{j)}^l \delta^{(3)}(\sigma, \sigma')$, and $\left\{\phi(\sigma), \pi(\sigma')\right\} = \delta^{(3)}(\sigma, \sigma')$ illustratively for generic matter fields, as well as the spatial diffeomorphism and Hamiltonian constraints $C_a =-2 \nabla_b p^{b} \mbox{}_a + C_a^{\text{matter}}$ and $H$. We choose a reference metric $\check q_{ij}$, which in its own adapted coordinates $\check a, \check b$ automatically satisfies $\check q_{\check r \check a} = \delta_{\check r \check a}$. We now impose the constraint $q_{\check r \check a} = \delta_{\check r \check a}$, which enforces that the metrics $q$ and $\check q$ differ at most in their (non-radial) $A,B$-components \cite{DuchObservablesForGeneral}. In order to show that $q_{\check r \check a} = \delta_{\check r \check a}$ is a good gauge fixing for $C_a$, we compute
\be
	\left\{ q_{\check r \check a}(\sigma),\ C_b[M^b] \right\} = 2 M_{(\check r; \check a)}(\sigma) \text{.} \label{eq:BracketQC}
\ee
Indeed, the vector field $M^b$ can always be chosen such that \eqref{eq:BracketQC} is non-vanishing, since the equation $2 M_{(\check r; \check a)} = \omega_{\check r \check a}$ is uniquely solvable for a given $\omega$ \cite{DuchObservablesForGeneral}. One can now pass to the Dirac bracket $\{\cdot,\cdot\}_{\text{DB}}$ implementing the constraints $C_b=0$ and $q_{\check r \check a} = \delta_{\check r \check a}$. The details of this procedure are spelled out in our longer companion paper \cite{BLSII}, since they are not essential for what follows. We find 
\begin{align}
	\hspace{-5mm} \left\{ q_{AB}(r, \theta), p^{CD}(r', \theta') \right\}_{\text{DB}} &= \delta_{(A}^C \delta_{B)}^D \delta(r, r') \delta^{(2)}(\theta, \theta') ~~~~~ ~~~~~~~~~\hspace{-20mm} \label{eq:Bracketqp} \\
	\left\{ q_{\check r \check a}(r, \theta), * \right\}_{\text{DB}} &= 0 \label{eq:BracketQ*}\\
	\left\{ F(r, \theta), p^{\check r \check a}(r', \theta') \right\}_{\text{DB}} &= -\mathcal L_{\underset{(r', \theta')}{\vec M}} F(r, \theta) \text{,} \label{eq:BracketFp}
\end{align}
where $*$ denotes an arbitrary phase space function, $F$ is any local function of $q_{AB}, p^{AB}, \phi, \pi$, and $\underset{(r', \theta')}{\vec M}$ is a vector field defined in \cite{BLSII}, and $\mathcal L$ is a Lie derivative acting on the internal space of $r,\theta$. We thus conclude that\\ 1) the $A,B$ components of the spatial metric and its momentum have canonical Dirac brackets, \\2) the constraint $q_{\check r \check a} = \delta_{\check r \check a}$ is consistent with \eqref{eq:BracketQ*}, and\\ 3) $p^{\check r \check a}$ acts via infinitesimal spatial diffeomorphisms. 

Passage to the reduced phase space, coordinatised by $q_{AB}$ and $p^{AB}$, now requires us to solve the spatial diffeomorphism constraint for $p^{\check r \check a}$ and insert the resulting expression in the Hamiltonian. The details are provided in \cite{BLSII}.\\

\noindent {\bf Quantisation}\\  
Starting from the Dirac brackets \eqref{eq:Bracketqp}, we can construct SU$(2)$ connection variables along standard lines, see e.g. \cite{BTTI} or our companion paper \cite{BLSIII} for details. We first extend the phase space to the canonical pair $\left\{K_A^i (r, \theta), E^B_j (r', \theta') \right\} = \delta(r,r')\delta^{(2)}(\theta,\theta') \delta_A^B \delta^i_j$ subject to the additional Gau{\ss} constraint $G_{ij} := E^{A}_{[i} K_{A|j]} = 0$, where $i,j = 1,2,3$ are now SU$(2)$ indices. Then, we perform a canonical transformation to the canonical pair $\left\{\m^{(\beta)}\!A_{A}^i(r, \theta),  \m^{(\beta)}\! E^B_j(r', \theta') \right\} = \delta(r,r')\delta^{(2)}(\theta,\theta') \delta_A^B \delta^i_j$, where $\m^{(\beta)}\!A^i_{A} = \frac{1}{2} \epsilon^{ijk} \Gamma_{Ajk} + \beta K_A^i$, $\Gamma_{Ajk}$ is the Peldan hybrid spin connection \cite{PeldanActionsForGravity}, and $\beta$ is a free parameter, similar to the Barbero-Immirzi parameter, and $\m^{(\beta)}\! E^A_i := E^A_i / \beta$. From these variables, we construct holonomies and fluxes as
\begin{align}
 	h_e(A) &:= \mathcal P \exp \left(- \int_e \m^{(\beta)}\! A_{Ai} \tau^i dx^A \right)   \\
	E_n(S) &:= \int_S \m^{(\beta)}\! E^{A}_i(\sigma) n^i(\sigma) \epsilon_{AB} \, dr \wedge dx^B \text{,}
\end{align}
where the $n^i$ are Lie algebra valued smearing functions and $\tau^i$ the Pauli matrices. We emphasise that the paths $e$ are tangential to the spheres $S^2_r \subset \Sigma$ of constant $r$, while the surfaces $S$ are foliated by radial geodesics. The reduced phase space, labelled by the variables $q_{AB}$ and $p^{CD}$, has thus been reexpressed via holonomies and fluxes with restricted path and surface labels. 
Up to this restriction, the corresponding holonomy-flux algebra is identical to the one from standard loop quantum gravity. Thus, quantisation can proceed along standard lines, see e.g. \cite{ThiemannModernCanonicalQuantum}, resulting in an $L^2$ space over the space of {\em restricted} generalised connections $\bar {\mathcal A}_{\text{res}}$, meaning with restricted paths as above. A generic element in this Hilbert space, a cylindrical function, will then depend on a finite number of holonomies, and consequently have support only at a finite number of radial distances $r$ from $\sigma_0$. A basis in the space of gauge invariant cylindrical functions is thus given by cylindrical functions depending on spin networks embedded in the spheres $S^2_r$. We will call such a basis element {\em multi spin network}.\\

\noindent {\bf Geometric operators}\\  
We will focus the discussion about quantum operators in this paper on the volume operator, since its construction highlights the main peculiarities coming from the radial gauge. The volume operator \cite{AshtekarDifferentialGeometryOn} plays a pivotal role in the definition of the LQG dynamics \cite{ThiemannQSD1}, where it enters through Poisson bracket identities, known as ``Thiemann's tricks''. We will employ a similar construction. Details of the construction of the Hamiltonian are provided in \cite{BLSIII}.

Since the delta-distribution in \eqref{eq:Bracketqp} is $3$-dimensional, we also need to define a $3$-volume operator in order to use an identity such as $e_A^i(\sigma) = \left\{A_A^i(\sigma), V_{\Delta} \right\} $, where $V_\Delta$ is the volume of a small open region containing $\sigma$. Classically, the volume of a region $R$ is given by $V(R) = \int_R\sqrt{q}  \, d^3x $ with 
\be
	\sqrt{q}  = \sqrt{V^i V_i}, ~~~~~  V^i := \frac{\beta^2}{2}  \epsilon^{ijk}  \m^{(\beta)}\!E^{A}_j  \m^{(\beta)}\!E^B_k \epsilon_{AB} \text{.}
\ee
We now try to promote this expression to an operator along the lines of \cite{AshtekarQuantumTheoryOf2, ThiemannQSD4}, that is we choose a lattice approximation of $R$, approximate the integrant by fluxes, and compute the action of the resulting operator in the limit of an infinitely refined lattice. 
A subtlety occurs when trying to remove the regulators of a given discretisation: while the integral provides three powers of the lattice spacing, two in the tangential directions and one in the radial direction, the fluxes under the integral absorb four powers, two in the tangential direction, and two in the radial direction. This means that we are left with an unused power of the radial lattice spacing. If this were a coordinate distance, the resulting operator would have the unacceptable property of being coordinate dependent. However, the radial distance is a physical distance, due to our chosen deparametrisation. The volume operator thus has a residual dependence on a {\em physical} regulator. We propose the following natural choice of a radial lattice to deal with the above problem:
we choose a set of real positive numbers $l_i$, $i \in \mathbb N_0$ with $l_0 = 0$, which define the extent of the radial smearing of fluxes (the dual lattice), and set $\Delta l_i := l_i - l_{i-1}$. Next, we choose real positive numbers $r_i$, $i \in \mathbb N$, such that $r_i \in (l_{i-1}, l_i)$, which correspond to lattice sites. We then approximate a radial integral as $\int_0^\infty dr f(r) \approx  \sum_{i=1}^{\infty} \Delta l_i f(r_i)$. We furthermore restrict our cylindrical functions to have support only at the lattice sites $r_i$. 

Following \cite{AshtekarQuantumTheoryOf2, ThiemannQSD4}, we then arrive at the volume operator
\begin{align}
	&\hat V(R) \Psi_\gamma   =  \sum_{v \in V(\gamma,R)} \hat V_v \, \Psi_\gamma  \label{eq:NewThreeVolumeOperator} \\
	&			\hat V_v \Psi_\gamma :=  \frac{\beta^2 \hbar^2}{8 \Delta l_i}\sqrt{\left( \sum_{e,e' \in E(\gamma,v)}  \, \text{sgn}(e, e') \epsilon^{ijk} \, R^e_j \, R^{e'}_k  \right)^2}  \Psi_\gamma    \text{,} \nonumber
\end{align}
where $\Psi_\gamma$ denotes a cylindrical function expressed on a graph $\gamma$ chosen such that at non-trivial (= not only parallel tangents) vertices all edges are ingoing, $V(\gamma,R)$ is the set of vertices of $\gamma$ contained in $R$, $E(\gamma,v)$ denotes the set of edges incident at the vertex $v$, $\text{sgn}(e, e')$ denotes the sign of $\epsilon_{AB} \dot e^A {{\dot{e}} {'}}^B$ at $v$, where $\dot e^A$, ${{\dot{e}} {'}}^B$ are the tangents of the edges $e,e'$, and $R^e_j$ are the standard right invariant vector fields. As expected, this operator is identical to the one of \cite{ThiemannQSD4} up to the factor of $\Delta l_i$.
The special properties of the volume operator in $2+1$ dimensions \cite{ThiemannQSD4} directly transfer to \eqref{eq:NewThreeVolumeOperator}. It generically does not vanish on two- or three-valent vertices, as long as there are at least two edges with non-parallel tangents.\\

\noindent {\bf Strategy for spherical symmetry} \\ 
\indent {\bf 1. Strategy:} Due to the coordinate system defined by the map $x^I \mapsto \exp_{\sigma_0} (x^I e^i_I)$, it is a natural approach to group-average the quantum states over rotations, defined in $\Sigma$ as the image of rotations in $T_{\sigma_0}\Sigma$ via the exponential map. Due to our choice of spherical coordinates, a rotation in $\Sigma$ just corresponds to changing the angular coordinates $(A,B) \equiv \theta$ in the same way at all radial coordinates. 
Due to the spherically symmetric setting, the coordinates $x^I$ have maximal range and span all of $\Sigma$ (up to non-trivial topology). This strategy, which is detailed in our companion paper \cite{BLSIII}, retains most degrees of freedom, as we will discuss later. Spherically symmetric cylindrical functions can now be defined by demanding invariance under such rotations, which have a straightforward action as moving holonomies in the quantum theory. They can be obtained by taking arbitrary cylindrical functions and group averaging them. The same applies to operators.  

{\bf 2. Strategy:} 
We recall \cite{KucharGeometrodynamicsOfThe} that the ADM formulation can be reduced to spherical symmetry via the ansatz $ds^2 = \Lambda^2(r,t) dr^2 + R^2(r,t) d\Omega^2$ for the spatial line element,
leading to the Poisson brackets $\{R(r),\ P_R(r')\} = \delta(r,r')$ and $\{\Lambda(r),\ P_{\Lambda}(r')\} = \delta(r,r')$. 
A necessary consequence of this reduction is that $p^{\check r \check A} = 0$, which already follows from the vanishing of spherically symmetric vector fields on $S^2$. Comparing with \eqref{eq:BracketFp}, we find that the generator of a certain class of spatial diffeomorphisms has to vanish. Furthermore restricting $a = A$ in \eqref{eq:BracketFp}, it follows that the involved vector fields span the set of vector fields tangential to spheres $S^2_r$, the image of the exponential map for radial coordinate $r$ \cite{BLSII}. Thus, $p^{\check r \check A} = 0$ can be imposed in the quantum theory by demanding invariance with respect to spatial diffeomorphisms which preserve the $S^2_r$. Technically, this is equivalent to the usual problem of implementing the spatial diffeomorphism constraint in loop quantum gravity \cite{AshtekarQuantizationOfDiffeomorphism}, however this time the conceptual difference is that we are averaging with respect to physical coordinates, which results in a reduction of physical degrees of freedom. The resulting quantum states live in the dual of the Hilbert space and consist essentially of diffeomorphism equivalence classes of spin networks lying on individual $S^2_r$.\\

\noindent {\bf Comparison of the two proposals} \\
\indent {\bf Degrees of freedom: } In the first strategy, the group averaging is performed with respect to the action of the compact group SO$(3)$. 
Specifying the edges of the spin networks thus retains an uncountable amount of information. The situation is different in the second strategy, which reduces spin networks to their corresponding diffeomorphism equivalence classes with respect to the physical coordinate system $(\theta)$, resulting in far less degrees of freedom \footnote{Which are countable up to $\theta$-moduli occurring for four and higher-valent vertices \cite{FairbairnSeparableHilbertSpace}}. The difference is essentially given by spherically symmetric correlations such as $\twopoint{\phi(r,\theta_1) \phi(r', \theta_2)} = \twopoint{\phi(r,g \theta_1) \phi(r', g \theta_2}$, where $\phi$ can represent e.g. a matter field or curvature scalar, and $g$ is a rotation. While the Hilbert space from the first strategy contains full information about such correlations, it is drastically reduced in the second proposal to diffeomorphism invariant correlations at $r=r'$.

{\bf Dynamics: } Quantum Hamiltonians (de\-pa\-ram\-e\-trised, as constraints, or master constraints) in the radial gauge can be constructed using slight extensions of the quantisation techniques of \cite{ThiemannQSD1, ThiemannQSD4, ThiemannQSD8}, as shown in \cite{BLSIII} \footnote{A slight technical issue at the point $\sigma_0$ is still under investigation.}. 
The regulators in the angular directions can always be removed, while integrals in the radial direction are approximated as before. For derivatives in the radial direction, we use standard finite difference approximations. Details are provided in \cite{BLSIII}.  
For angle-independent lapse, the Hamiltonians preserve the spherically symmetric states from the first strategy, which ultimately results from using quantisation techniques for diffeomorphism invariant theories for the angular directions. However, the diffeomorphisms preserving the $S^2_r$ from the second strategy do not commute with such Hamiltonians, since they do not map radial geodesics into radial geodesics with respect to the original metric, interfering with the reduced phase space structure of the Hamiltonian, which contains non-local contributions from integrals along the radial direction \cite{BLSII}. 

A different strategy is thus necessary for the second approach, based on operators which are invariant with respect to $S^2_r$-preserving diffeomorphisms. Our proposal for such Hamiltonians consists in defining the spherically symmetric midisuperspace Hamiltonians, e.g. the one of \cite{KucharGeometrodynamicsOfThe}, as operators on the full theory Hilbert space. The radial gauge corresponds to setting $\Lambda = 1$ and solving for $P_{\Lambda}$ in terms of $R$ and $P_R$ via the spatial diffeomorphism constraint, see \cite{BLSII} for details. We then have to represent only $R$ and $P_R$ as operators, which can be easily done again using standard techniques \cite{ThiemannQSD1, ThiemannQSD4}. Since both $R(r)$ and $P_R(r)$ can be obtained from averages of functions of the unreduced variables $q_{AB}$ and $p^{AB}$ over $S^2_r$, they are naturally invariant with respect to $S^2_r$-preserving diffeomorphisms. Hamiltonians constructed along the above lines thus preserve the spherically symmetric states of the second strategy. This strategy is suited best for a true Hamiltonian resulting from deparametrisation, since no issues related to a proper reduction of degrees of freedom at the quantum level occur.

In both strategies, the issue of anomalies in the Hamiltonian constraint is so far unresolved. This problem can however be improved upon by using non-rotating dust to deparametrise the Hamiltonian constraint \cite{BrownDustAsStandard, HusainTimeAndA, SwiezewskiOnTheProperties}, leading to a single true Hamiltonian, and thus the absence of the anomaly issue. For pure general relativity, a suitable deparametrisation of the Hamiltonian constraint, leading to a manageable true Hamiltonian, is not known so far. In this case, we would resort to the Master constraint approach \cite{ThiemannQSD8}.\\

\noindent {\bf Choice of spin networks and comparison to other approaches} \\
A natural question to ask now is the relation of the presented reduction strategies to previous work based on quantising classically reduced models, such as \cite{KastrupCanonicalQuantizationOf, KucharGeometrodynamicsOfThe, GambiniLoopQuantizationOf}. Here, the choice of spin networks in our proposal is crucial: in the radial gauge, a spherically symmetric midisuperspace model (coupled to additional matter) has $R(r)$ and $P_R(r)$ as gravitational degrees of freedom. Our spin networks at the radial lattice points $r_i$ however generically encode more degrees of freedom than just the total area $A_{S^2_r} = 4 \pi R^2$. Interestingly, due to the peculiar properties of the $2+1$-dimensional volume operator \cite{ThiemannQSD4}, there exists a choice of spin network encoding exactly one degree of freedom, being a Wilson loop with exactly one kink, labelled only by the SU$(2)$ spin $j$. If we restrict ourselves only to such Wilson loops, one for each lattice site $r_i$, we obtain a maximally simple model, for which closed and manageable formulas for the action of the Hamiltonians from both strategies can be obtained, since the volume operator is diagonal and non-trivial on the kink and we choose a graph-preserving regularisation for the field strength of the connection. The corresponding object in a midisuperspace quantisation is then the quantum number associated to $R^2$ (e.g. $k_i$ associated to $\hat E^x$ in \cite{GambiniLoopQuantizationOf}) at a given lattice site.
A comparison of the explicit dynamics of our models to a midisuperspace quantisation using the radial gauge has not been performed so far and will be left for further research. 

Another interesting aspect of the dynamics, which has received much attention in recent years, is the issue of coarse graining and choice of appropriate spin networks to describe physical situations, see \cite{DittrichTheContinuumLimit} and references therein. In our model, this issue reappears in the choice of spin networks for the spheres $S^2_r$. While the total volume can be encoded in a Wilson loop as above, it is unclear how the dynamics would be affected by choosing a more refined spin network encoding the same total volume, but more local information. Due to the simplified dynamics (the volume operator is still diagonal on three-valent vertices), this issue can be investigated using spin networks based on graphs which are triangulations of the $S^2_r$.\\

\noindent {\bf Conclusion} \\ 
We have outlined the construction of a computable framework to study spherically symmetric quantum gravity dynamics. The key classical ingredient has been a geometric deparametrisation of the spatial diffeomorphism constraint, leading to a family of effectively $2+1$-dimensional holonomy-flux algebras. The quantum dynamics becomes tractable because the spin network vertices for the quantum states can be chosen to be two- or three-valent due to the special properties of the $2+1$-dimensional volume operator. This work opens the possibility to study spherically symmetric quantum dynamics within full loop quantum gravity, such as dust collapse in the Lema\^{i}tre-Tolman-Bondi model. \\


\noindent {\bf Acknowledgements}\\
This work was partially supported by the grant of Polish Narodowe Centrum Nauki nr 2011/02/A/ST2/00300 and by the grant of Polish Narodowe Centrum Nauki nr 2013/09/N/ST2/04299.
NB was supported by a Feodor Lynen Research Fellowship of the Alexander von Humboldt-Foundation and gratefully acknowledges discussions with Antonia Zipfel.

\end{document}